# Structure and dynamics of $Fe_{90}Si_3O_7$ liquids close to Earth's liquid core conditions

*Ling Tang,[1] Chao Zhang,[2] Yang Sun,[3,7] Kai-Ming Ho,[3] Renata M. Wentzcovitch,[5,6,7] and Cai-Zhuang Wang[3,4*]*

[1] *Department of Applied Physics, College of Science, Zhejiang University of Technology, Hangzhou 310023, China*

[2] *Department of Physics, Yantai University, Yantai 264005, China*

[3] *Department of Physics and Astronomy, Iowa State University, Ames, Iowa 50011, USA*

[4] *Ames Laboratory-USDOE, Iowa State University, Ames, Iowa 50011, USA*

[5] *Department of Earth and Environmental Sciences, Columbia University, New York, New York 10027, United States*

[6] *Lamont–Doherty Earth Observatory, Columbia University, Palisades, New York 10964, United States*

[7] *Department of Applied Physics and Applied Mathematics, Columbia University, New York, New York 10027, United States*

**Abstract**

Using an artificial neural-network machine learning interatomic potential, we have performed molecular dynamics simulations to study the structure and dynamics of $Fe_{90}Si_3O_7$ liquid close to the Earth's liquid core conditions. The simulation results reveal that the short-range structural order (SRO) in the $Fe_{90}Si_3O_7$ liquid is very strong. About 80% of the atoms are arranged in crystalline-like SRO motifs. In particular, ~70% of Fe-centered clusters can be classified as either hexagonal-close-pack (HCP/HCP-like) or icosahedral (ICO/ICO-like) SRO motifs. The SRO clusters centered on Fe, Si, or O atoms are strongly intermixed and homogenously distributed throughout the liquid. The atomic structure of the liquid and the fractions of dominant SRO clusters are not sensitive to pressure/temperature used in the simulations except that the SRO of the O-centered clusters is enhanced close to inner core pressures. The O diffusion coefficient is about 2-3 times larger than the Fe and Si ions and increases more rapidly in the deeper core regions.

## 1. Introduction

The Earth's outer core is believed to be composed of a liquid iron alloy with up to 10% of light elements such as silicon, oxygen, sulfur, carbon, or hydrogen [1, 2]. Despite extensive studies, this region's chemical composition and temperature are still elusive and controversial [3]. Owing to the ultrahigh pressures (> 135 GPa) and temperatures (> 3800 K), experimental and theoretical studies at core conditions are limited. However, the first experiment beyond core conditions was successfully carried out in 2010 [4]. Molecular dynamics (MD) simulation [5, 6] has been a versatile computational tool for investigating condensed systems' structure and dynamic properties with applications in condensed matter physics, materials science, chemistry, biology, engineering, and earth sciences. However, accurate and efficient descriptions of interatomic forces are crucial to perform reliable MD simulations for such a complex system and under extreme environments of Earth's outer core.

Recently, we have developed an artificial neural network machine learning (ANN-ML) interatomic potential for Fe-Si-O system [7]. We showed that the developed ANN-ML interatomic potential describes well the structure and dynamics of the Fe-Si-O system at high pressures (> 100 GPa) and high temperatures (> 3000 K). The aim was to enable accurate MD simulations of materials containing these three elements at extreme high-pressure and temperature conditions. The achieved accuracy of the developed ANN-ML interatomic potential allows us to investigate at an atomistic level the structure and dynamics of this complex Fe-Si-O system in the Earth's liquid core.

In this paper, we systematically investigate the structure and dynamics of Fe-Si-O liquids with the composition $Fe_{90}Si_3O_7$ (~1.6 wt% Si and 2.1 wt% O) at different pressures/depths of the Earth's liquid core using this potential and MD simulations. This composition is close to the estimated one in the Earth's outer core [8-10]. Pressure-temperature (P-T) conditions at different depths are modeled assuming a linear relationship between pressure (ranging from 135-363 GPa) and temperature (ranging from 3800-6500K) with depth in the Earth's liquid core. Pressures larger than 323 GPa may represent the outer core in Earth's early days when the inner core was smaller and hotter. Short-range order (SRO) analyses of MD trajectories enable us to reveal the structure and dynamics of the liquid at a quantitative level.

The paper is organized as follows. In Section 2, we describe the details of the simulation and analysis methods. The simulation results are presented and discussed in Section 3. We give our summary in Section 4.

## 2. Simulation and Analysis Methods

MD simulations of liquid $Fe_{90}Si_3O_7$ are performed using the previously developed ANN-ML interatomic potential [7, 11] and the LAMMPS package [12]. The MD simulation cell is a cubic box containing 10,000 atoms (9,000 Fe, 300 Si, and 700 O atoms) and periodic boundary conditions along all three directions. We carried out six simulations at 135 < P < 363 GPa and 3800 < T < 6500K, assuming a linear P-T relationship. These P-T conditions are 135GPa/3800K, 181GPa/4340K, 226GPa/4880K, 272GPa/5420K, 317GPa/5960K, and 363GPa/6500K respectively. An isothermal-isobaric (NPT) ensemble at a given pressure and a Nose-Hoover thermostat [13, 14] at a given temperature are used in the simulations. The MD time step is 2.5 fs. The initial liquid structure is prepared by melting the cubic lattice at ultrahigh-temperature T=10000K and high pressure 210 GPa, then by adjusting the simulation box size at 6500K using NVT ensemble to obtain a liquid at pressure/temperature of 363GPa/6500K. The liquid structures at other pressures/temperatures are prepared by cooling from the 363GPa/6500K using NPT simulation. At each pressure/temperature, NPT MD simulations are performed for 25ps to thermalize the sample at the given pressure/temperature, followed by 25ps of simulation to acquire the trajectories for structure analysis and by 1ns of simulation to calculate dynamic properties for a sufficient statistical average.

*SRO structure analysis* – Liquid structures are characterized by pair correlation functions (PCF), *g*(*r*). PCF is a conditional probability density of finding a particle at distance *r*, given that there is a particle at the coordinate origin. Thus *g*(*r*) provides a measure of local spatial ordering in a liquid. Mathematically, partial PCF between the atom type $\alpha$ and $\beta$ is given by $g_{\alpha\beta}(r) = \rho_{\alpha\beta}^{-2}\langle\sum_i \sum_{j\neq i} \delta(\vec{r}_{i\alpha})\delta(\vec{r}_{j\beta} - r)\rangle$, where $\rho_{\alpha\beta} = \rho_0\sqrt{a_\alpha a_\beta}$ is the corresponding partial density with $\rho_0$ being the atomic density of the liquid and $a_\alpha$ and $a_\beta$ the atomic concentrations of the corresponding elements [15].

To gain more insight into the local SRO structure motifs in the liquid at different pressure/temperatures, the liquid samples generated by the MD simulations are also analyzed by the cluster alignment method previously developed in [16]. Quasi-spherical clusters formed by a centered atom and 30 neighbor atoms are extracted from the liquid samples. We refer to these atomic clusters as sample clusters. The sample clusters are first aligned against several known templates, including HCP or BCC etc. This process is referred to as template alignment, which classifies sample clusters to commonly observed template motifs. The alignment score *s* measures the similarity and is defined as,

$$s = \min_{0.8\leq\alpha\leq 1.2}\left\{\left[\frac{1}{N}\sum_{i=1}^{N}\frac{(\vec{r}_{ic}-\alpha\vec{r}_{it})^2}{(\alpha\vec{r}_{it})^2}\right]^{1/2}\right\} \quad (1)$$

where N is the number of atoms in the template. $\vec{r}_{ic}$ and $\vec{r}_{it}$ are the atomic positions in the sample cluster and template, respectively. $\alpha$ is chosen between 0.8 and 1.2 to vary the template size for an optimal alignment. The smaller score indicates the higher similarity between the aligned cluster and the template. The simulated annealing algorithm is employed to minimize the alignment scores. The internal structures of both the sample and template clusters are fixed. Only rigid translation and rotation of the whole cluster are allowed during the minimization. The bond lengths of the template are allowed to scale by a factor of α to fit the sample cluster better. The clusters are classified as the template motif with which it shows the lowest alignment scores. If the scores are larger than 0.16, the cluster represents a new motif that cannot be classified to any known template. To analyze SRO in these unclassified clusters, we perform a pairwise alignment [17] to compute the similarity matrix between any two sample clustesrs. The clique analysis [18] is employed to organize similar clusters into cliques. By superposing the clusters from the same clique, one can identify new SRO motifs from the distribution of the first shell atoms.

*Dynamical property analysis* - To quantitatively study the dynamic properties, we calculated the self-diffusion constants $D$ of every element in the Fe-Si-O liquids. The mean-square displacement (MSD) as a function of time is given by [15, 19]

$$\langle R_\alpha^2(t) \rangle = \frac{1}{N_\alpha} \langle \sum_{i=1}^{N_\alpha} |R_{i\alpha}(t+\tau) - R_{i\alpha}(\tau)|^2 \rangle, \quad (2)$$

where $N_\alpha$ is the total atomic number of α species, $R_{i\alpha}$ are coordinates of atom $i$ and $\tau$ is the arbitrary time origin. The MSD of the liquids in the limit of a long time should behave linearly with the time, and the slope of the line gives the self-diffusion constant $D$ by the Einstein relationship,

$$D = \lim_{t \to \infty} \langle R_{i\alpha}^2(t) \rangle / 6t \quad (3)$$

## 3. Results and Discussions

Fig. 1 shows the partial pair-correlation functions (PCFs) of $Fe_{90}Si_3O_7$ liquid from 135GPa/3800K to 363GPa/6500K. The MD trajectories over 25ps are collected to calculate the PCFs. As pressure/temperature increases, the position of the first peak shifts to a smaller $r$, as intuitively expected. It is interesting to note that in O-O PCF, as pressure/temperature increases, the height of the first peak decrease while the size of the second peak increases, suggesting the O atoms in the first shell move outwards to the second shell as pressure increases. The figures on the right also show pressure dependence of averaged coordination number for Fe-, Si-, and O-center atoms in $Fe_{90}Si_3O_7$ liquid, where the first minimum of partial PCFs is used as cutoff distance to calculate the coordination number. It clearly shows the O-O average coordination number decreases from 1.3 to 0.6, while the average coordination number of the other atom pairs increases as pressure increases. More details about the partial and total coordination number distributions around each element in the liquids at the two extreme pressure/temperature conditions can also be seen in Fig. 2. We can see that although the total coordination of O ions changes very little with pressure, the O-Fe coordination increases while the coordination number of O-O pairs within the first shell decreases with increasing pressure, which is consistent with the trend of O-O partial PCFs obtained above. We also note that the coordination number distributions around the Fe and Si atoms in the liquid are not sensitive to pressure.

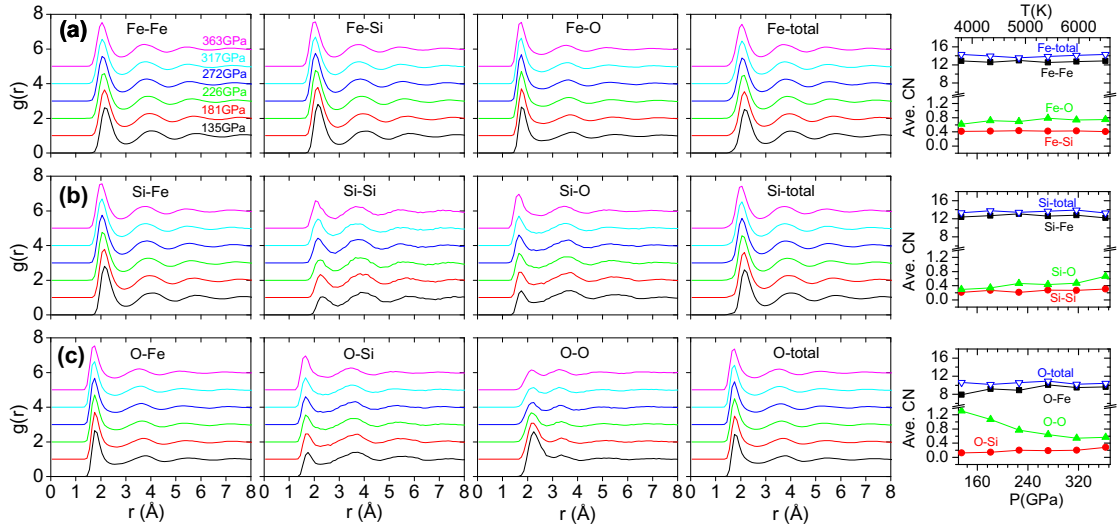

Fig. 1. The partial pair-correlation functions of $Fe_{90}Si_3O_7$ liquid at different pressures and temperatures. The pressure/temperature dependences of averaged coordination number for atoms in $Fe_{90}Si_3O_7$ liquid is also shown on the right of the figure.

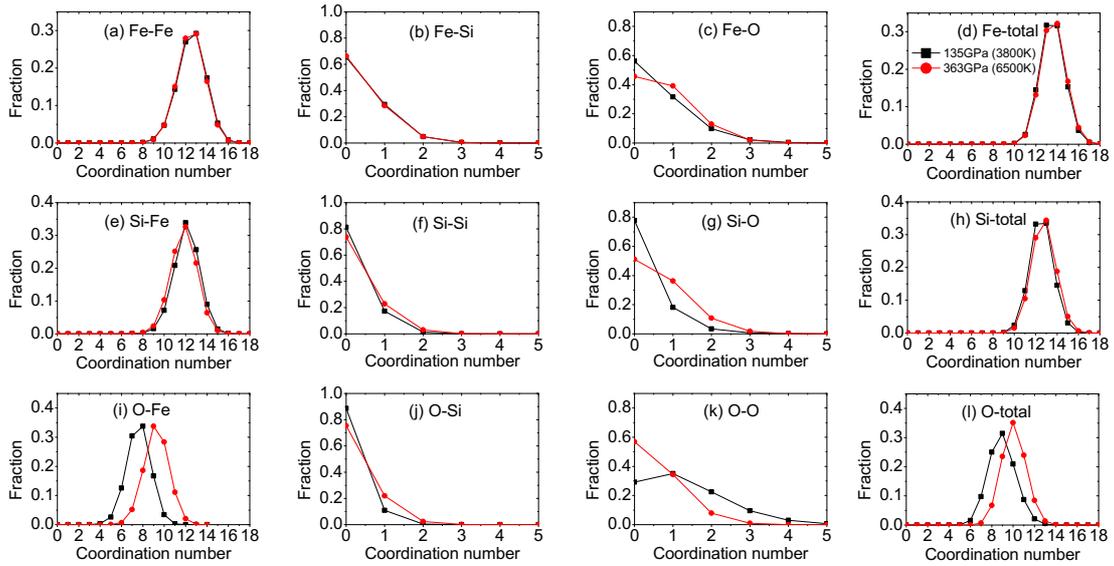

Fig. 2. Coordination number distributions for $Fe_{90}Si_3O_7$ liquid at 135GPa/3800K and 363GPa/6500K.

To gain more insight into the SRO motifs in the liquid structures, we first use the template alignment method to analyze the structures of $Fe_{90}Si_3O_7$ liquids from our MD simulations. Six commonly observed SRO motifs (i.e., icosahedra (ICO), Octahedral (OCT), tetrahedral (TET), face-center-cubic (FCC), body-center-cubic (BCC), and hexagonal close-packed (HCP)) are used as templates for the alignment analysis. The Fe-, Si-, and O-centered clusters with a total of 30 neighboring atoms are extracted from the inherent structure of the liquid samples. After the alignment, a cutoff value of 0.16 (noted that the smaller score indicates the aligned cluster and template are more similar to each other) is used to assign the clusters to the corresponding template motifs. We find that only about 10% of the clusters in the liquid samples can be assigned to these given templates. Therefore, these commonly observed SRO motifs are inadequate to describe the liquids' SRO.

To investigate the possibility of other SROs motifs in $Fe_{90}Si_3O_7$ liquids, pairwise cluster alignments [17] plus clique analysis [18] were performed iteratively to identify new SRO motifs in the liquid samples. Such alignment plus analysis can group similar structures and find their common motifs without knowing the motifs in liquid samples in advance [17]. We use the liquid sample at 317GPa/5960K for such motif search. First, we exclude all the clusters identified by the template alignment discussed above from the liquid sample. Then the remaining unclassified clusters are aligned with each other. If the alignment score between the two clusters is less than 0.2, these two clusters are considered to be similar. After the pairwise alignments among all clusters are done, the clique analysis algorithm is used to identify the maximum clique of similar clusters. The common structure of these similar clusters is then used as a new template (which is referred to as PW1) and added to the known templates list to perform the new-round of template cluster alignment for all the clusters in the sample. SRO clusters identified by the new-round of template alignment are then excluded from liquid sample. Next, the new iteration of pairwise alignment and clique analysis are performed to identify new possible SRO motif (which is referred to as PW2). Then the PW2 motif is addded to the templates list for next iteration of template alignment, and so on, until about 80% clusters in the sample are identified as some types of SRO.

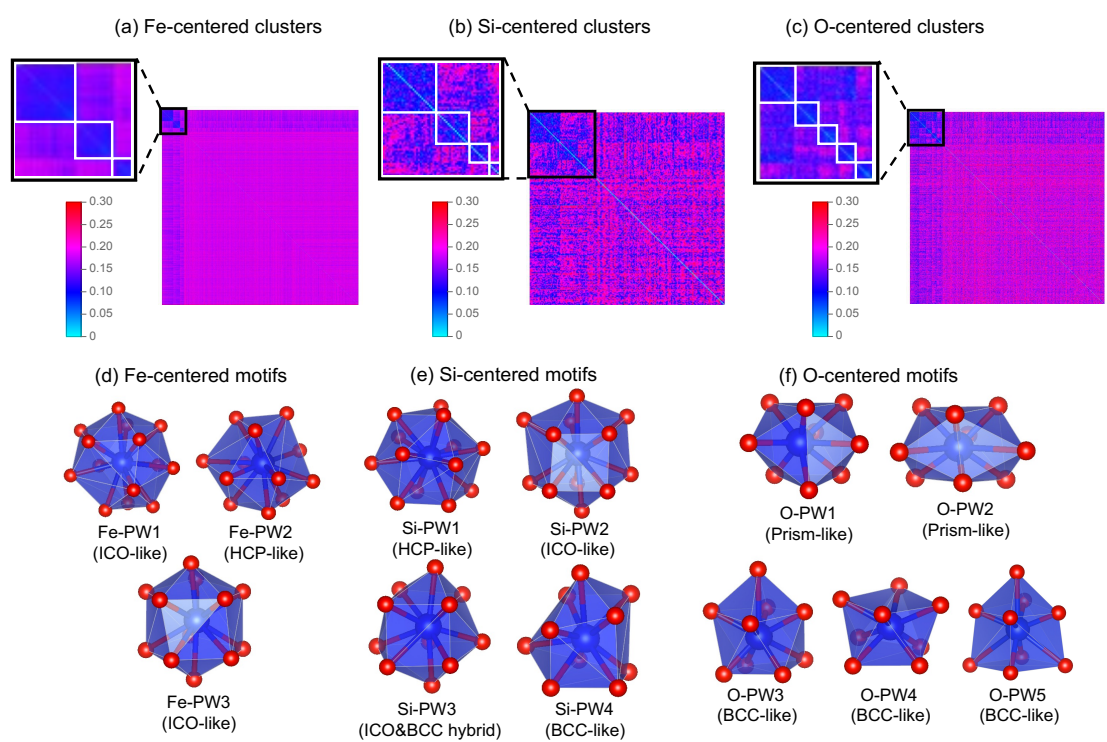

Fig. 3. The alignment scores matrix plot for pairwise (a) Fe- (b) Si- and (c) O-centered clusters, where the conventional SRO clusters are excluded. The obtained dominant motifs for (d) Fe- (e) Si and (f) O-centered clusters are also plotted.

Fig. 3(a)-(c) shows the matrix plot of pairwise alignment scores sorted by the clique analysis for Fe-, Si-, and O-centered clusters, where the six commonly observed SRO clusters identified by the template alignment are excluded. Three, four, and five iterations of pairwise alignment plus clique analysis are performed for the Fe-, Si-, and O-centered clusters, respectively. Thus, we found new Fe-PW1 to 3, Si-PW1 to 4, and O-PW1 to 5 motifs of the Fe-, Si-, and O-centered clusters, respectively. The structures of obtained SROs motifs are plotted in Fig. 3(d)-(f), respectively. The pairwise alignment scores of these similar clusters form the blue diagonal square in Fig. 3(a)-(c). The obtained pairwise motifs exhibit some distortions from the ideal lattices and can be classified as HCP-like, ICO-like, BCC-like, and Prism-like motifs. For Fe-centered clusters, Fe-PW2 is HCP-like, while Fe-PW1 and Fe-PW3 are ICO-like SROs. For Si-centered clusters, Si-PW1 is HCP-like while Si-PW2 is ICO-like SROs, and Si-PW4 is BCC-like SROs. Si-PW3 is hybrid motifs of ICO and BCC, which has half ICO and half BCC structure. For O-centered clusters, O-PW1 and O-PW2 are Prism-like SROs. O-PW3, O-PW4, and O-PW5 are BCC-like SROs.

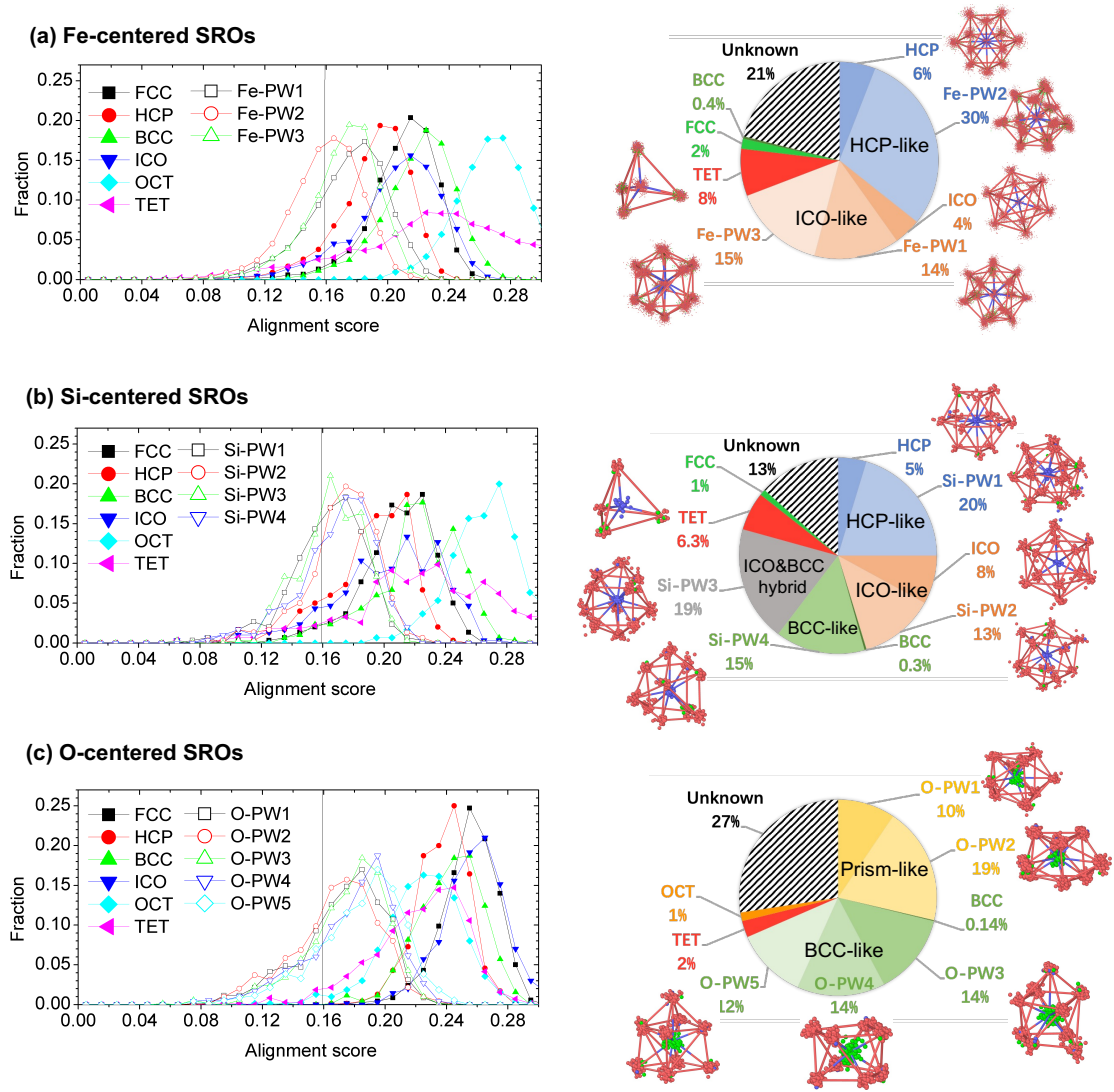

Fig. 4. The alignment score distributions of conventional SROs and pairwise alignment found SROs motifs for (a) Fe- (b) Si and (c) O-centered clusters. The populations of various motifs obtained by counting the number of the clusters with alignment scores below 0.16 are shown on the right.

Using the new SRO motifs revealed by the pairwise alignment shown in Fig. 3 and the six conventional motifs discussed above as templates, comprehensive template-cluster alignments have been performed for the liquid samples. Fig. 4 shows the alignment score distributions of all SROs motifs found in $Fe_{90}Si_3O_7$ liquid at 317GPa/5960K. The populations of various motifs below the cutoff score of 0.16 are also plotted in the right column of Fig. 4. We can see that the alignments with the new templates from Fig. 3 give lower score than those from the conventional templates, indicating the clusters in the liquid are more similar to the new templates. One can also see that ICO/ICO-like and HCP/HCP-like motifs are the dominant SROs for Fe- and Si-centered clusters in the sample, with a fraction of 69% and 46%, respectively. The BCC-like motif is the dominant SRO in O-centered SROs (40%) and has a significant fraction in Si-centered clusters (15%). Moreover, the sample shows a substantial fraction (29%) of the Prism-like SRO motif for the O-centered clusters and 19% ICO&BCC hybrid SRO motif for Si-centered clusters. A noticeable fraction (8.0% and 6.3%, respectively) of TET clusters are also observed for the Fe-centered and Si-centered clusters, respectively, in the sample. In Fig. 4, the ionic positions of the clusters in the major SRO motifs are plotted after the alignment. We can see that the ion distribution in each motif is well localized around the vertexes of the corresponding template, indicating that our cluster alignment method is accurate for classifying the SRO motifs of the clusters in the liquid.

To see if the new SRO motifs obtained from the pairwise alignment and clique analysis are sensitive to pressures, we repeat the same analysis for the sample at 181GPa/4340K and obtain the same set of new templates. Using these new templates and the six conventional templates, we perform the template-cluster alignment to classify the fractions of different SRO motifs in all six liquid samples obtained from our MD simulations. The distributions of the alignment scores for the other five samples are all similar to that of the sample at 317GPa/5960K shown in Fig. 4. More details of the fractions of the major SRO motifs as the function of pressure/temperature in the pressure range 135-363GPa are shown in Fig. 5. We can see the variations of the SROs for Fe-centered and Si-centered clusters as the function of pressure/temperature are not significant. Some variations in the Si-centered clusters would also be attributed to the poor statistics because Si concentration is only 3% in the sample. By contrast, the Prism-like and BCC-like SROs for the O-centered clusters are found to increase systematically with increasing pressure, as seen in Fig. 5 (c).

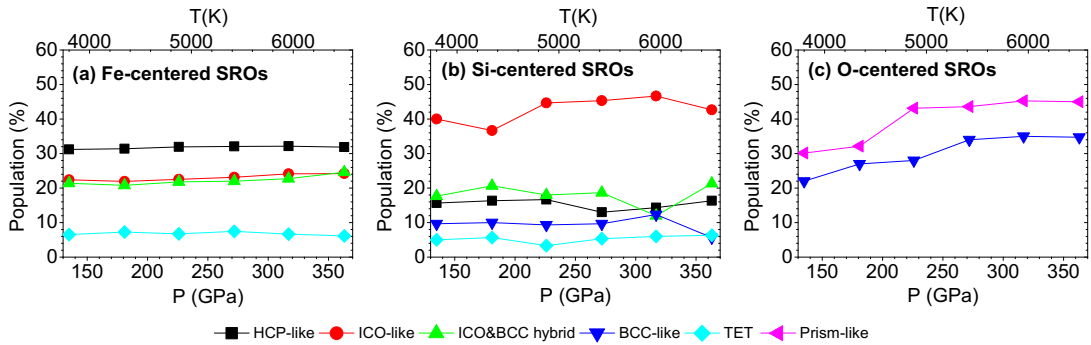

Fig. 5. The pressure/temperature dependence of fractions of dominant SROs in $Fe_{90}Si_3O_7$ liquid, where all the dominant SROs are classified as HCP-like, ICO-like, ICO&BCC hybrid, BCC-like, Prism-like, and TET motifs.

The spatial distribution of the dominant Fe-, Si- and O-centered SROs, reprinting by the center atoms with different colors, in the liquid samples at 181GPa/4340K and 317 GPa/5960K are shown in Fig. 6 (a), (b), and (c) respectively. We see that these SROs clusters distribute quite homogenously and inter-mixed each other throughout the simulation cell. Although some degree of clustering of the same SRO can be seen for all types of SRO motifs, no apparent phase separations are observed.

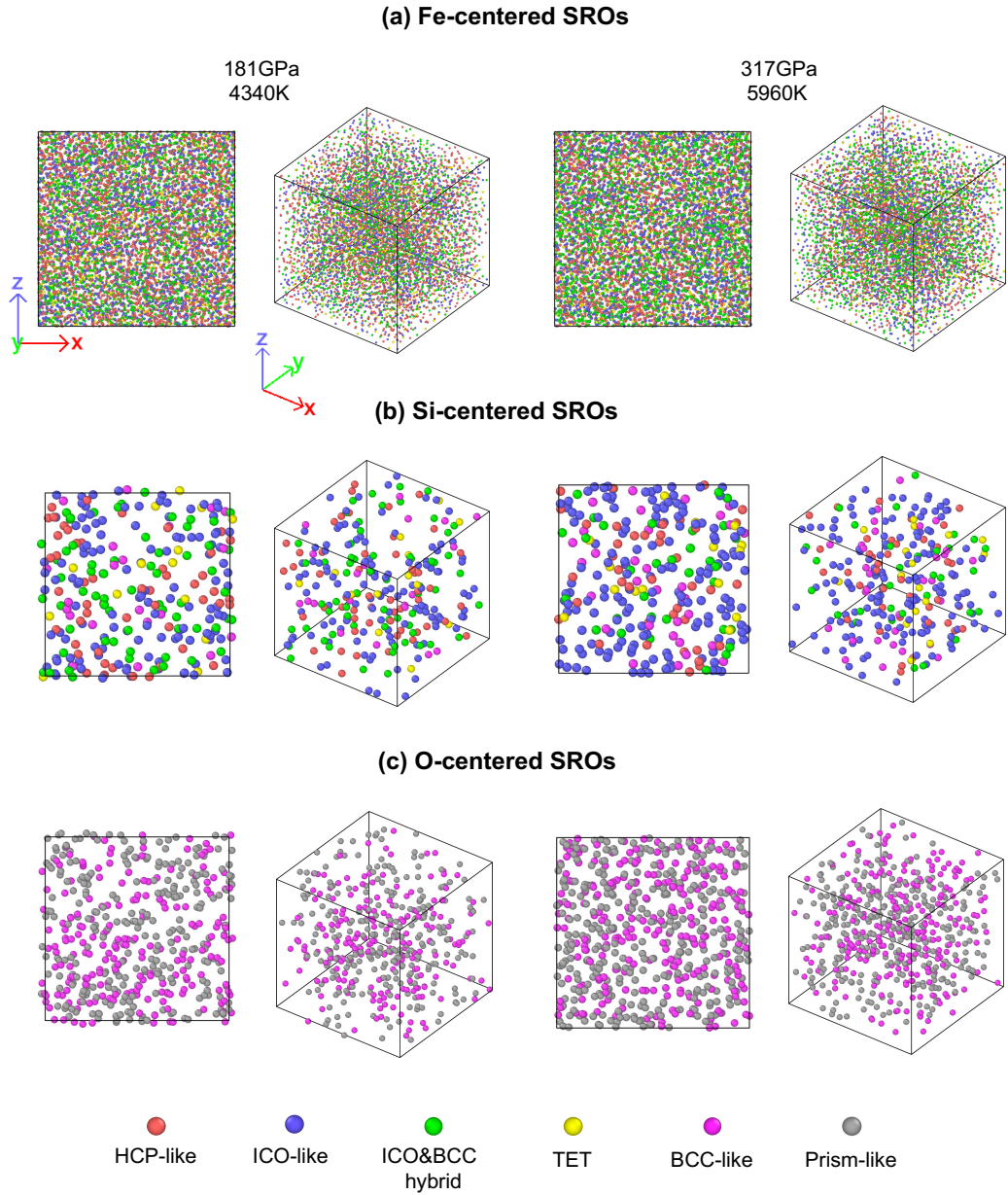

Fig. 6. The spatial distribution of (a) Fe- (b) Si- and (c) O-center atoms of dominant SROs in the low pressure (181GPa) and high pressure (317GPa) sample.

Finally, Fig. 7 shows the pressure/temperature dependence of Fe, Si, and O self-diffusion constants in the liquid from 135GPa/3800K to 363GPa/6500K. Fe and Si self-diffusion constants are similar and increase linearly with pressure. The O self-diffusion constant is about 2-3 times greater than Fe or Si. The rate of self-diffusion constants increase for O is similar to Fe or Si for pressure < 320GPa. At pressures > 320GPa, the self-diffusion constant of O increases more rapidly than Fe or Si atoms. The significant changes in the O coordination number shown in Fig. 1 and Fig. 2 and the fractions of the SRO motifs in the O-centered clusters shown in Fig. 5 would be attributed to the faster diffusion of O atoms and the dependence of the diffusion on pressure/temperature.

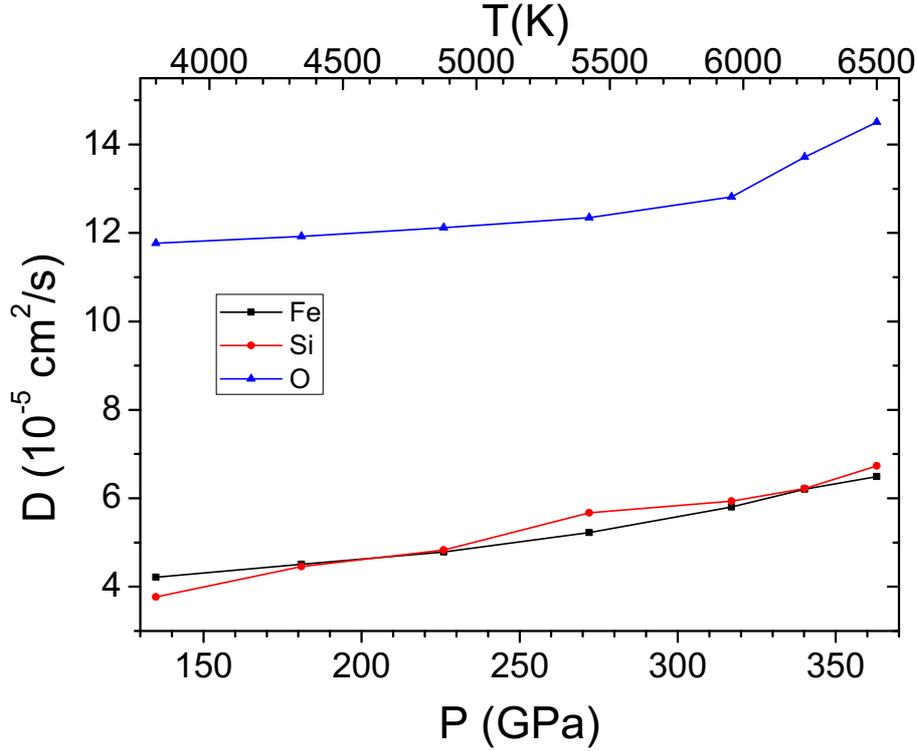

Fig. 7. The pressure (temperature) dependency of self-diffusion constants of Fe, Si, O in the liquid from 135GPa/3800K to 363GPa/6500K.

## 4. Summary

Using an accurate artificial neural-network machine learning (ANN-ML) interatomic potential, we have performed extensive MD simulations to systematically investigate the structures and dynamics of a $Fe_{90}Si_3O_7$ liquid at high pressures and temperatures resembling those of the Earth's liquid core. The SRO structure in the liquid is analyzed using cluster alignment methods based on atomic coordinates produced by the MD simulations. The simulation and analysis reveal strong structural SRO in the $Fe_{90}Si_3O_7$ liquid under the Earth's liquid core conditions. More than 2/3 of Fe atoms in the liquid can be classified as the center atoms for either HCP/HCP-like or ICO/ICO-like first shell clusters, while about 2/3 of O-centered clusters exhibit BCC-like or Prism-like SRO motifs. The SRO clusters around the minority Si atoms are also strong and span over four different SRO motifs (i.e., HCP-like, BCC-like, ICO-like, and ICO&BCC hybrid) with a population of about 15-20% each. These dominant SRO clusters are strongly intermixed and homogenously distributed throughout the samples. The atomic structures for the liquids at different depths of Earth's liquid core are similar, except for some enhancement of the SROs around O in the deep core regions. While the diffusion constant of Fe and Si atoms are comparable and increase almost linearly with depth in the Earth's liquid core, the O diffusion constant is about 2-3 times larger than that of the Fe and Si and increases more rapidly in the deeper core regions.

It should be noted that the chemical composition and the pressure/temperature profile used in our simulations are close to but not the same as those in the Earth's outer core [3]. Therefore, caution is advised when interpreting our simulation results in the context of the Earth's outer core. Nevertheless, our simulation results show that the overall SRO motifs in the liquid are very robust and not sensitive to the pressure or temperature in the investigated pressure/temperature range.

**Acknowledgements**
Work at Ames Laboratory was supported by the U.S. Department of Energy (DOE), Office of Science, Basic Energy Sciences, Materials Science and Engineering Division including a grant of computer time at the National Energy Research Supercomputing Center (NERSC) in Berkeley. Ames Laboratory is operated for the U.S. DOE by Iowa State University under contract # DE-AC02-07CH11358. Work at Iowa State University and Columbia University was supported by the National Science Foundation awards EAR-1918134 and EAR-1918126. L. Tang acknowledges the support by the National Natural Science Foundation of China (Grant No. 11304279). C. Zhang was supported by the National Natural Science Foundation of China (Grants No. 11874318）.

**Data availability**
The data that support the findings of this study are available from the corresponding author upon reasonable request.

**Conflict of interest**
The authors declare no competing interests.